\documentclass[aps, prb, preprint, superscriptaddress, onecolumn]{revtex4-1}

\usepackage{amsmath}
\usepackage{amsfonts}
\usepackage{amssymb}
\usepackage{graphicx}
\usepackage{multirow}
\usepackage{xcolor}

\begin{document}

\title{Photoinduced structural dynamics of multiferroic TbMnO$_3$}
\date{\today}

\author{Elsa Abreu}
\email{elsabreu@phys.ethz.ch}
\affiliation{Institute for Quantum Electronics, Eidgen\"ossische Technische Hochschule (ETH) Z\"urich, 8093 Z\"urich, Switzerland}

\author{Matteo Savoini}
\affiliation{Institute for Quantum Electronics, Eidgen\"ossische Technische Hochschule (ETH) Z\"urich, 8093 Z\"urich, Switzerland}

\author{Larissa Boie}
\affiliation{Institute for Quantum Electronics, Eidgen\"ossische Technische Hochschule (ETH) Z\"urich, 8093 Z\"urich, Switzerland}

\author{Paul Beaud}
\affiliation{Paul Scherrer Institut, 5232 Villigen, Switzerland}

\author{Vincent Esposito}
\affiliation{Paul Scherrer Institut, 5232 Villigen, Switzerland}

\author{Martin Kubli}
\affiliation{Institute for Quantum Electronics, Eidgen\"ossische Technische Hochschule (ETH) Z\"urich, 8093 Z\"urich, Switzerland}

\author{Martin J. Neugebauer}
\affiliation{Institute for Quantum Electronics, Eidgen\"ossische Technische Hochschule (ETH) Z\"urich, 8093 Z\"urich, Switzerland}

\author{Michael Porer}
\affiliation{Paul Scherrer Institut, 5232 Villigen, Switzerland}

\author{Urs Staub}
\affiliation{Paul Scherrer Institut, 5232 Villigen, Switzerland}

\author{Bulat Burganov}
\affiliation{Institute for Quantum Electronics, Eidgen\"ossische Technische Hochschule (ETH) Z\"urich, 8093 Z\"urich, Switzerland}

\author{Chris Dornes}
\affiliation{Institute for Quantum Electronics, Eidgen\"ossische Technische Hochschule (ETH) Z\"urich, 8093 Z\"urich, Switzerland}

\author{Angel Rodriguez-Fernandez}
\affiliation{Paul Scherrer Institut, 5232 Villigen, Switzerland}

\author{Lucas Huber}
\affiliation{Institute for Quantum Electronics, Eidgen\"ossische Technische Hochschule (ETH) Z\"urich, 8093 Z\"urich, Switzerland}

\author{Gabriel Lantz}
\affiliation{Institute for Quantum Electronics, Eidgen\"ossische Technische Hochschule (ETH) Z\"urich, 8093 Z\"urich, Switzerland}

\author{Jos\'e R. L. Mardegan}
\affiliation{Paul Scherrer Institut, 5232 Villigen, Switzerland}

\author{Sergii Parchenko	}
\affiliation{Paul Scherrer Institut, 5232 Villigen, Switzerland}

\author{Jochen Rittmann}
\affiliation{Paul Scherrer Institut, 5232 Villigen, Switzerland}

\author{Cris Svetina}
\affiliation{Paul Scherrer Institut, 5232 Villigen, Switzerland}

\author{Gerhard Ingold}
\affiliation{Paul Scherrer Institut, 5232 Villigen, Switzerland}

\author{Steven L. Johnson}
\email{johnson@phys.ethz.ch}
\affiliation{Institute for Quantum Electronics, Eidgen\"ossische Technische Hochschule (ETH) Z\"urich, 8093 Z\"urich, Switzerland}
\affiliation{Paul Scherrer Institut, 5232 Villigen, Switzerland}

\begin{abstract}
We use time-resolved hard x-ray diffraction to investigate the structural dynamics of the multiferroic insulator TbMnO$_3$ in the low temperature antiferromagnetic and ferroelectrically ordered phase.
The lattice response following photoexcitation at 1.55 eV is detected by measuring the (0 2 4) and (1 3 -5) Bragg reflections.
A 0.02$\%$ tensile strain, normal to the surface, is seen to arise within 20 - 30 ps.
The magnitude of this transient strain is over an order of magnitude lower than that predicted from laser-induced heating, which we attribute to a bottleneck in the energy transfer between the electronic and lattice subsystems.
The timescale for the transient expansion is consistent with that of previously reported demagnetization dynamics.
We discuss a possible relationship between structural and demagnetization dynamics in TbMnO$_3$, in which photoinduced atomic motion modulates the exchange interaction, leading to a destruction of the magnetic order in the system.

DOI: 
\end{abstract}

\maketitle

\newpage


Rare-earth manganites are prime examples of complex materials, where a rich phase diagram is driven by a subtle interplay of charge, spin, orbital and structural degrees of freedom.
Identifying the role of each degree of freedom in the determination of key physical properties in such materials is essential but often extremely challenging.
One way to address this challenge is to use pump-probe measurements to explore how different degrees of freedom respond to being driven out of equilibrium.

Here we focus on insulating TbMnO$_3$, a prototypical multiferroic with an orthorhombically distorted perovskite structure, which exhibits two magnetic phase transitions, at $T_{N_1} \simeq 42$ K and $T_{N_2} \simeq 27$ K, driven by ordering of the Mn $3d$ spins \cite{Kimura2003}.
Below $T_{N_1}$, TbMnO$_3$ is ferromagnetic along the $a$-axis, antiferromagnetic along $c$, and it exhibits an antiferromagnetic sinusoidal spin density wave order along $b$.
Below $T_{N_2}$, a magnetic cycloid forms in the $bc$-plane which gives rise to a ferroelectric polarization along the $c$-axis.
Pump-probe experiments in the multiferroic phase ($T < T_{N_2}$) \cite{Handayani2013, Kubacka2014, Johnson2015, Bowlan2016, Bothschafter2017, Baldini2018} have shown that the antiferromagnetic order decreases on a $\sim$20 - 30 ps time scale in response to strong photoexcitation at photon energies of 1.55 eV and 3 eV, which directly excite an intersite Mn $3d$ transition and an O $2p$ $\rightarrow$ Mn $3d$ transition, respectively \cite{Bastjan2008, Moskvin2010}.
A phenomenological model \cite{Johnson2015} has associated this decrease in magnetic order to an increase in effective spin temperature, up to $\sim$45 K for an absorbed fluence of 9 mJ/cm$^2$.
It is, however, still unclear exactly how the electronic transitions excited by the pump pulse couple to the magnetism; several different hypotheses have been suggested \cite{Talbayev2015, Johnson2015, Bothschafter2017, Baldini2018}.
In this work, we investigate the structural dynamics of TbMnO$_3$ using ultrafast hard x-ray diffraction in an attempt to investigate a potential contribution of the lattice to this process.\\


The experiments were performed at the FEMTO slicing beamline of the Swiss Light Source at the Paul Scherrer Institute \cite{Beaud2007}, using a gated two dimensional Pilatus detector \cite{Henrich2009}.
Bulk [010]-cut samples of TbMnO$_3$ were cooled down to approximately 20 K using a liquid helium jet.
The temperature at the sample position was calibrated in advance and monitored throughout the experiment with the help of a temperature sensor placed next to the sample.
To prevent freezing of water or nitrogen on the sample, a box was built around it and purged with helium gas.
The p-polarized 1.55 eV pump beam and the monochromatic 7.05 keV x-ray probe beam were incident from the same direction but with different grazing incidence angles on the sample of 5.5$^\circ$ (or 10.2$^\circ$, in a subsequent experiment) and 0.5$^\circ$, respectively.
The intensity profile of the pump beam has a penetration depth (1/e), normal to the surface, of 323 nm for a polarization along the crystallographic $a$-axis and 671 nm for a polarization along the crystallographic $c$-axis \cite{Baldini2018}.
The electric field profile of the x-ray probe has a penetration depth (1/e) of 76 nm \cite{Henke1993}.
The pumped area is about four times larger than the probed area.
The pump and probe pulses have full-width-at-half-maximum durations of about 100 fs and 120 fs and repetition rates of 1 kHz and 2 kHz, respectively.
The temporal overlap between the two pulses was determined by time-resolved x-ray diffraction measurements performed on bulk [411]-cut bismuth samples at the start of the experiment, for calibration \cite{Johnson2008}.\\

\begin{figure} [htb!]
\begin{center}
\includegraphics[width=0.5\textwidth,keepaspectratio=true]{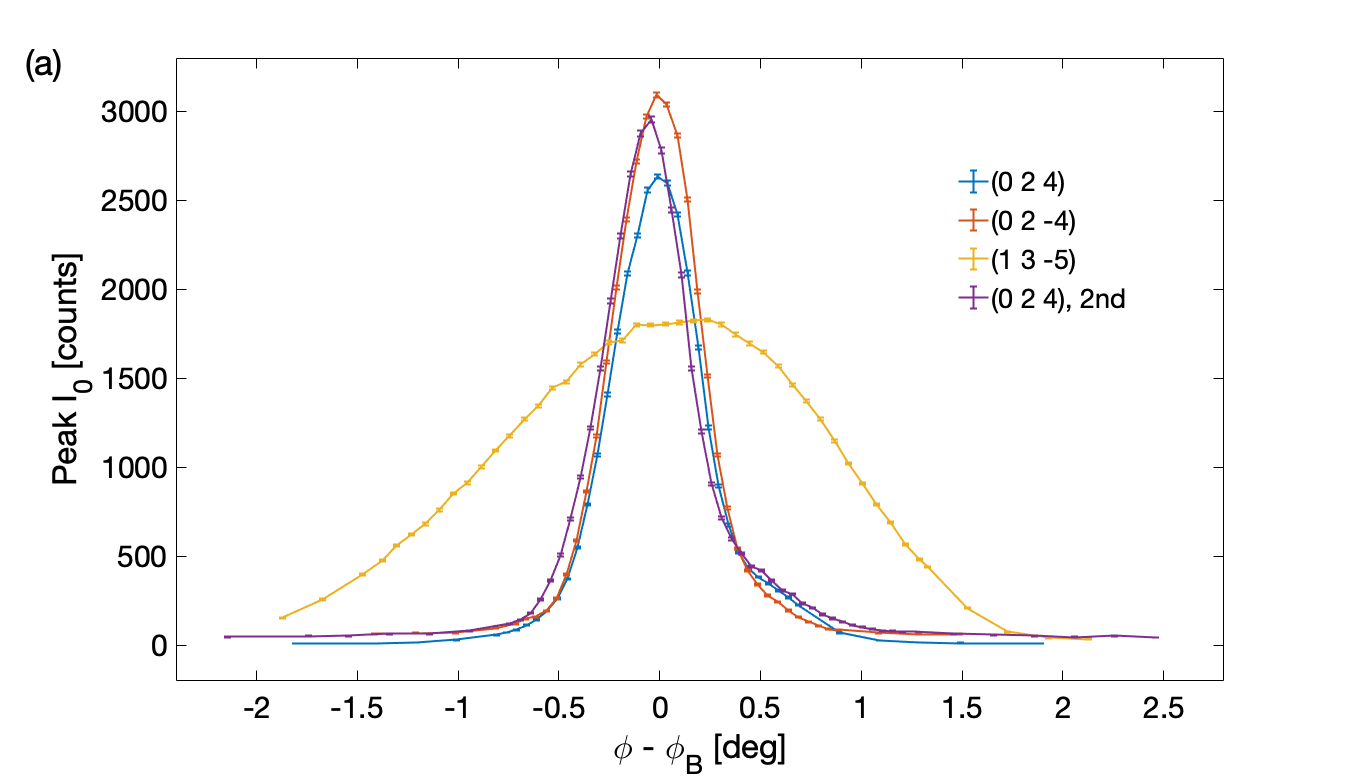}
\includegraphics[width=0.5\textwidth,keepaspectratio=true]{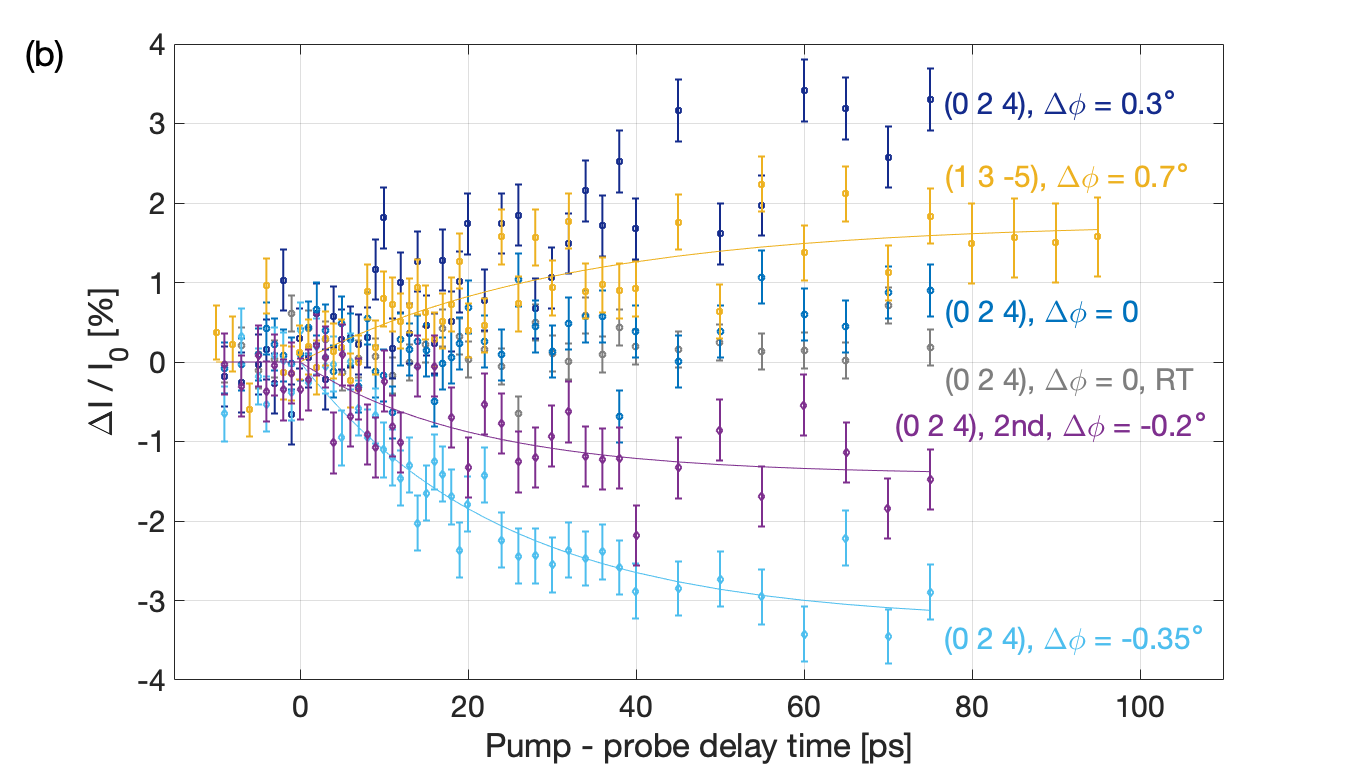}
\caption{
(a) Equilibrium $\phi$-scan of the relevant Bragg reflections, at $T$ = 20 K.
(b) Pump - probe delay traces at fixed $\Delta\phi = \phi - \phi_B$ positions and for different Bragg reflections, as labeled.
All data were taken at $T$ = 20 K except for the trace labeled RT (room temperature).
The (1 3 -5) trace was acquired at a pump fluence of 4.5 mJ/cm$^2$, the others at 9 mJ/cm$^2$.
Full lines are fits to an exponential function, as described in the text.
}
\label{fig1}
\end{center}
\end{figure}

Figure \ref{fig1}a shows the variation in intensity $I_0$ of four lattice reflections as a function of $\phi-\phi_B$, where $\phi$ is the angle of rotation of the sample around the normal to its surface and $\phi_B$ is the value of $\phi$ that maximizes a particular diffraction peak.
Here, the intensity $I_0$ is taken as the number of photons detected during one second in a region containing the diffraction peak, for a particular value of $\phi$.
Data is shown for the (0 2 4) and (1 3 -5) Bragg reflections, as well as for (0 2 -4).
Note that for the low temperature phase of TbMnO$_3$ (space group \emph{Pbnm}), (0 2 -4) is equivalent to (0 2 4) and exhibits the same dynamics in this [010]-cut sample geometry, so that data from the two reflections are equivalent.
A second $\phi$-scan for (0 2 4), measured during a subsequent experiment, is also included in Fig. \ref{fig1}a as a reproducibility check.
A small background intensity is visible in some $\phi$-scans, due to incoherent light scattering onto the detector.
This contribution is, however, essentially independent of $\phi$ and therefore has no impact on the results and analysis presented below.

We investigate the structural dynamics of TbMnO$_3$ by measuring the intensity diffracted from the (0 2 4) and (1 3 -5) planes following photoexcitation.
Given the scattering geometry dictated by these reflections at grazing incidence, the projection of the p-polarized 1.55 eV pump wavevector onto the sample surface forms 49$^\circ$ and 29$^\circ$ angles with respect to the crystallographic $c$-axis for the (0 2 4) and (1 3 -5) Bragg reflections, respectively.

Two types of time-resolved measurements were performed.
In the first type, the value of $\phi$ is kept fixed and only the pump - probe delay is varied.
Time traces obtained in this way for absorbed fluences of 4.5 and 9 mJ/cm$^2$ are presented in Fig. \ref{fig1}b for different $\phi$ values, Bragg reflections and temperatures ($T$ = 20 K and room temperature, RT).
The data are shown as the change in peak intensity following photoexcitation, $\Delta I(t) = I(t) - I_0$, relative to the unpumped case, $I_0$.
No significant photoinduced effect is detected at the peak of the $\phi$-scan, for $\Delta\phi = \phi-\phi_B=0$.
Traces taken on the right (left) side of the $\phi$-scan, for $\phi-\phi_B > 0$ ($\phi-\phi_B < 0$), do show an increase (decrease) in $\Delta I(t) / I_0$, consistent with a shift of the peaks towards larger $\phi$ values.
Fits were performed to a function of the form
\begin{equation}
f(t) = A \: H(t) \: [1 - \text{exp}(-t/\tau)],
\label{fitfun}
\end{equation}
where $H(t)$ is the Heaviside function, and $A$ and $\tau$ are fit parameters.
The resulting fits are included in Fig. \ref{fig1}b for the three traces for which $\tau$ can be determined within a meaningful uncertainty such that $\tau > 2 \, \delta\tau$, where $\delta\tau$ is the standard error.
The fitted $\tau$ values are shown in Table \ref{table}.

\begin{table}
\centering
\begin{tabular}{cccc}
\textbf{Reflection} & \textbf{Fluence} & $\boldsymbol{\phi-\phi_B}$ & $\boldsymbol{\tau}$ \\
 & \textbf{(mJ/cm$^2$)} & & \textbf{(ps)} \\
\hline
\hline
(0 2 4) & 9 & -0.35$^\circ$ & 24$\pm$3 \\
(0 2 4), 2nd & 9 & -0.2$^\circ$ & 21$\pm$8 \\
(1 3 -5) & 4.5 & 0.7$^\circ$ & 31$\pm$10 \\
\hline
(0 2 4) & 9 & - & 30$\pm$10 \\
(0 2 4), 2nd & 9 & - & 18$\pm$5 \\
(0 2 -4) & 4.5 & - & 19$\pm$21 \\
(1 3 -5) & 4.5 & - & 40$\pm$16 \\

\end{tabular}
\caption{
Fit results for the data in Fig. \ref{fig1}b (top) and Fig. \ref{fig2} (bottom).
}
\label{table}
\end{table}

\begin{figure} [htb!]
\begin{center}
\includegraphics[width=0.5\textwidth,keepaspectratio=true]{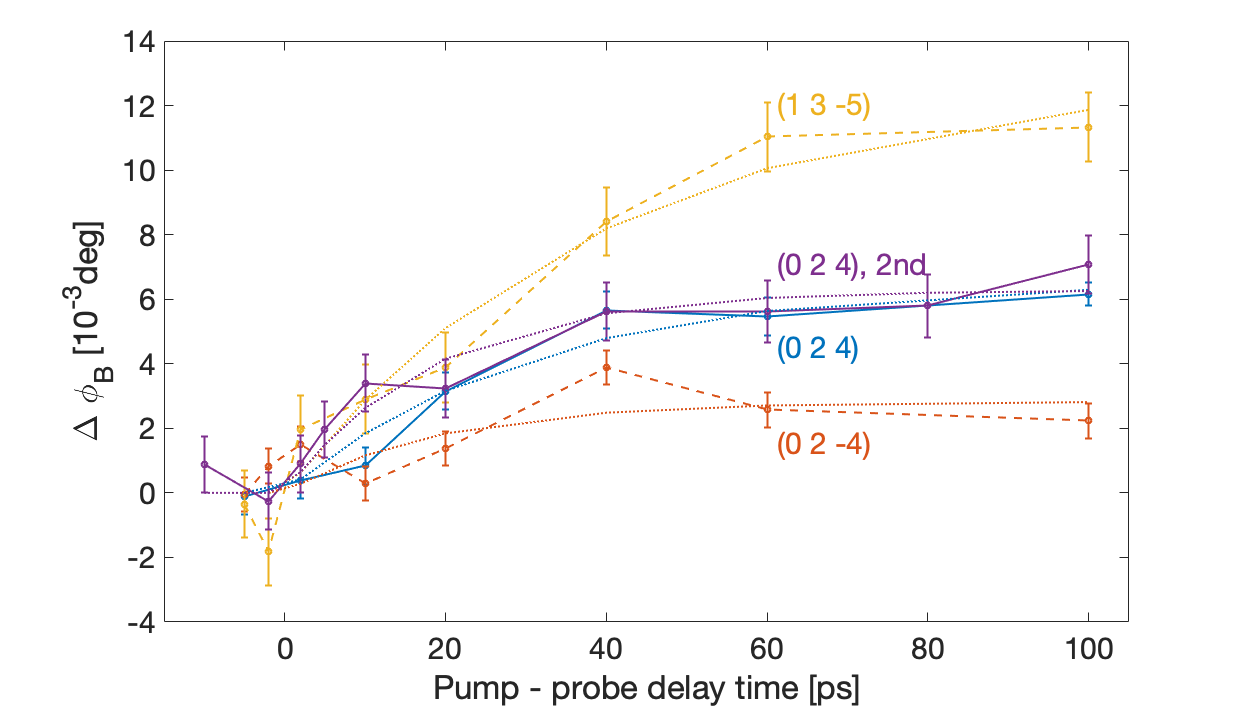}
\caption{
Photoinduced $\phi$ shift for different Bragg reflections, at $T$ = 20 K.
Full (dashed) lines correspond to an absorbed fluence of 9 mJ/cm$^2$ (4.5 mJ/cm$^2$).
The error bars are estimated by bootstrapping.
Dotted lines are fits to an exponential function, as described in the text.
}
\label{fig2}
\end{center}
\end{figure}

In the second type of measurements, full $\phi$-scans are measured at different pump - probe delays for the Bragg reflections shown in Fig. \ref{fig1}, and for absorbed fluences of 4.5 and 9 mJ/cm$^2$.
This approach provides a more complete picture of the structural evolution of the system, since it enables a distinction between changes in the $\phi$-integral of the Bragg peak intensity and shifts of the Bragg reflection along $\phi$.
The $\phi$-integral of the peak intensity is seen to remain unaffected by photoexcitation up to a pump-probe delay of 100 ps within the resolution of our measurement, which can detect changes larger than $\sim 1\%$.
A shift in the peak position along $\phi$ is, however, observed, as shown in Fig. \ref{fig2}.
Each data point shown in Fig. \ref{fig2} is calculated from the first moment of the $\phi$-scans
\begin{equation}
\Delta\phi_B = \frac{\int (\phi-\phi_0) I(\phi) d\phi}{\int I(\phi) d\phi}
\end{equation}
where $\phi_0$ is chosen such that $\Delta\phi_B = 0$ for the unpumped sample, and the integration is carried out over the range of the scan in $\phi$ (Fig. \ref{fig1}a).
As seen in Fig. \ref{fig1}a, the (1 3 -5) $\phi$-scan does not perfectly return to the baseline on the $\phi - \phi_B < 0$ side.
We estimate that the values of $\Delta \phi_B$ for the (1 3 -5) Bragg reflection presented in Fig. \ref{fig2} could be overestimated but by no more than $\sim 10^{-3} \: \text{deg}$, which is within the noise level of the measurements.
\footnote{The estimate was done by calculating $\Delta \phi_B$ with even more points removed on the $\phi - \phi_B < 0$  side.
Removing up to four points reduced $\Delta \phi_B$(100 ps) by less than $10^{-3} \: \text{deg}$ compared to the unaltered $\phi$-scan shown in Fig. \ref{fig1}a, which is within the noise level of the measurements.
Removing up to four points on the $\phi - \phi_B > 0$ side had no effect on the calculated $\Delta \phi_B$ value.}
From Fig. \ref{fig2} we see that $\Delta \phi_B$ is positive, as anticipated from the time traces in Fig. \ref{fig1}b.
$\Delta \phi_B$ values vary approximately linearly with fluence (comparing the (0 2 4) and (0 2 -4) data sets), and differ between Bragg reflections, as expected.
Fig. \ref{fig2} also shows fits of each data set to Eq. \ref{fitfun}.
The fit parameters, all reported in Table \ref{table}, yield timescales similar to those extracted from Fig. \ref{fig1}b, although with overall larger uncertainties.
It is clear from the data of Fig. \ref{fig2} and from the fit results that the peak shift dynamics approach saturation within our 100 ps measurement window.

A conversion from $\Delta I(t) / I_0$ (Fig. \ref{fig1}b) to $\Delta \phi_B$ can also in principle be done, but it relies on an accurate determination of the derivative of the $\phi$-scan at the $\phi$ position chosen for the $\Delta I(t) / I_0$ measurement, as well as on the assumption that there is no change in the shape of the $\phi$-scan.
We therefore opt to extract $\Delta \phi_B$ for Fig. \ref{fig2} from the full $\phi$-scan analysis, and use the data in Fig. \ref{fig1}b to obtain the timescales shown in Table \ref{table}.\\

\begin{figure} [htb!]
\begin{center}
\includegraphics[width=0.5\textwidth,keepaspectratio=true]{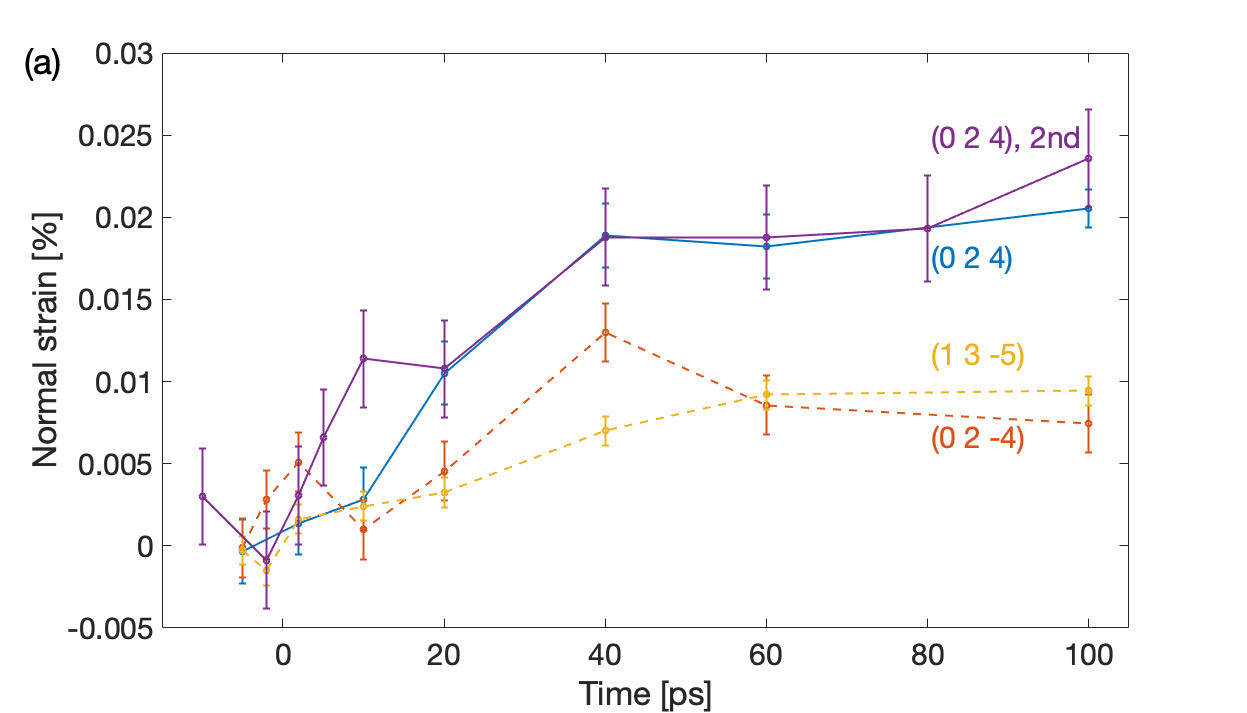}
\includegraphics[width=0.5\textwidth,keepaspectratio=true]{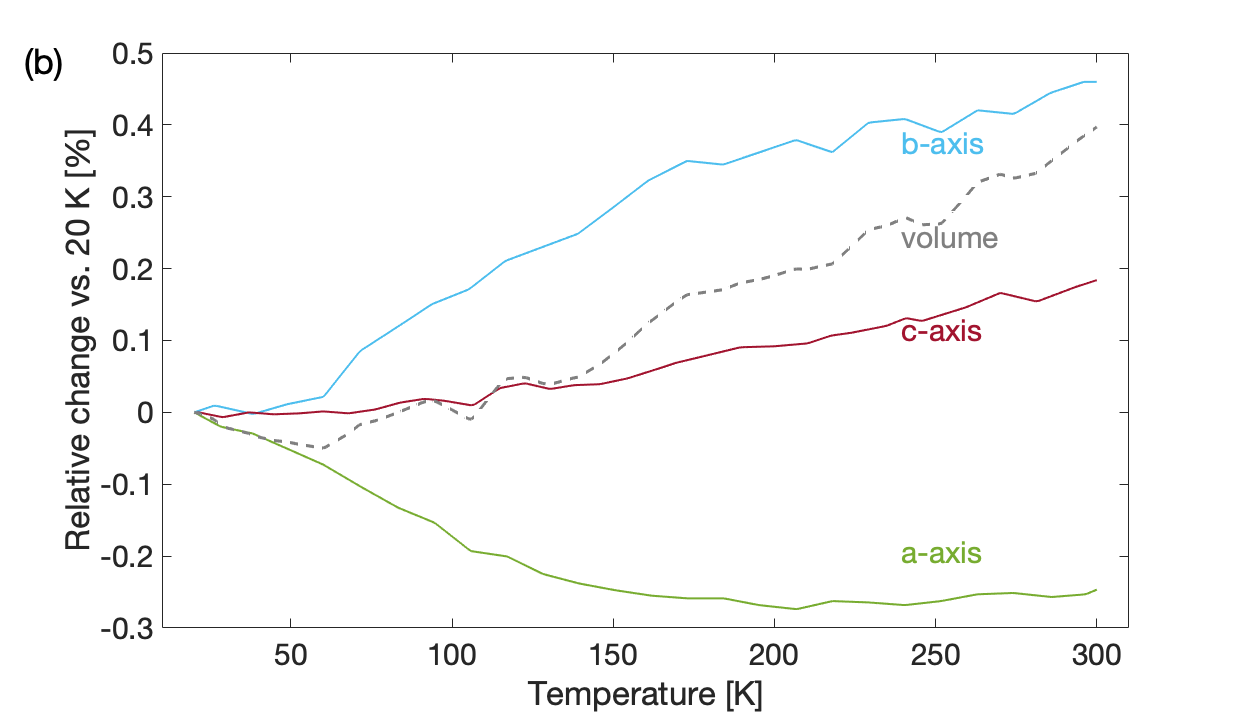}
\caption{
(a) Photoinduced strain normal to the sample (along the $b$-axis), estimated from Fig. \ref{fig2} as described in the text.
Full (dashed) lines correspond to an absorbed fluence of 9 mJ/cm$^2$ (4.5 mJ/cm$^2$).
(b) Axial strains and volume change as a function of temperature for bulk TbMnO$_3$ (from Blasco \emph{et al.} \cite{Blasco2000}), relative to $T$ = 20 K.
}
\label{fig3}
\end{center}
\end{figure}

We now discuss the significance of our results in understanding the response of multiferroic TbMnO$_3$ to strong photoexcitation.
1.55 eV photons promote intersite $d-d$ transitions between Mn$^{3+}$ ions, creating Mn$^{2+}$ - Mn$^{4+}$ pairs \cite{Bastjan2008, Moskvin2010}.
This electronic energy is then partially transferred to the lattice in the form of heat.
One consequence of lattice heating is an increase in structural disorder, which would manifest in diffraction as an increase in the Debye-Waller factor that suppresses the intensity of diffraction peaks.
This would appear in our measurements as a drop in the integrated intensity of our $\phi$-scans which, as discussed above, was not observed.
We conclude that this effect is small, both at 20 K and at RT, compared to our experimental uncertainties and given our choice of Bragg reflections (choosing larger reciprocal lattice vectors could enhance it).
Consequently, our experiment is not able to directly measure the timescale of this structural disorder due to lattice heating.
In lieu of a direct experimental measurement of the lattice heating, we can provisionally assume a typical timescale for this process on the order of one picosecond \cite{Wall2009, Baldini2018}.
On sufficiently long time scales, the lattice eventually thermalizes at a temperature which can be estimated from the specific heat \cite{Choithrani2011} and the absorbed fluence.
In our measurements on TbMnO$_3$, for an absorbed fluence of 9 mJ/cm$^2$ and an initial $T$ = 20 K
\footnote{We analyzed the effect of average heating by comparing a $\phi$-scan on the (1 3 -5) Bragg reflection when the pump laser was off with one where it was on, arriving after the probe pulse and leading to an absorbed fluence of 4.5 mJ/cm$^2$.
We estimate an effective initial temperature of at most 70 K instead of the nominal value of 20 K.
The estimate was done by converting between $\Delta \phi_B$ and $\eta_b$ as described below, and between $\eta_b$ and $T$ using Fig. \ref{fig3}b).
While this observation changes the absolute temperature values in our discussion, the discrepancy we report between the measured $\eta_b(100$ ps) and the expected $b$-strain value remains, as does the fact that a ten times lower absorbed fluence would be required to explain our data.
}, we estimate a temperature increase of $\Delta T \sim$ 157 K ($\Delta T \sim$ 136 K), up to a final temperature $T_f \sim$ 177 K ($T_f \sim$ 156 K) for measurements performed on the (0 2 4) ((1 3 -5)) reflection.
The dependence of $\Delta T$ on the reflection is related to the strong in-plane anisotropy \cite{Bastjan2008, Baldini2018} of the [010]-cut TbMnO$_3$ samples.
\footnote{The projection of the pump electric field along each in-plane crystallographic axis, which differs for the (0 2 4) and (1 3 -5) Bragg reflections, as specified above, was used to determine the ratio of the absorbed fluence along the $a$- and $c$-axis, the change in reflection along both axes being negligible.
The deposited energy density along each axis was then calculated as the ratio between absorbed fluence and penetration depth, and the total deposited energy density as a weighted sum of those two contributions.}

Heating the lattice by $\Delta T$ has the effect of creating expansive thermal stress near the surface of the crystal, which leads to the development of expansive uniaxial strain \cite{Thomsen1986}.
This expansion can be related to the $\Delta \phi_B$ value extracted from our data.
In an isotropic material, the uniaxial strain induced by the thermal stress is normal to the surface.
Although TbMnO$_3$ is not isotropic, the sample surface is along a principal axis of the elastic tensor \cite{Hazama2000, Choithrani2011}.
This implies that the strain should also be normal to the surface in our case.
Studies of similar manganite systems do not take into account the anisotropy in the elastic properties \cite{Choithrani2009, Choithrani2011} or find it to be small \cite{Hazama2000, Lalitha2012}.
We will therefore assume that the photoinduced strain is normal to the sample surface, i.e. along the \emph{b}-axis.
Under this assumption, $\Delta \phi_B$ can be calculated for the (0 2 4) and (1 3 -5) Bragg reflections assuming different values of the normal component of the strain, $\eta_b$.
\footnote{In practice, we calculate $\phi_B$ starting from the low temperature structure \cite{Blasco2000} but for different values of the lattice parameter along the \emph{b}-axis, corresponding to the different strain values.}
The relationship between $\Delta \phi_B$ and $\eta_b$ is verified to be approximately linear, and the extracted proportionality constant is used to convert the $\Delta \phi_B(t)$ values shown in Fig. \ref{fig2} to $\eta_b(t)$ values, shown in Fig. \ref{fig3}a.
As expected, the differences observed between Bragg reflections in $\Delta \phi_B(t)$ are no longer present in $\eta_b(t)$.
The only observable dependence is on fluence, with a larger fluence leading to a larger $\eta_b(t)$, as expected.

Strain arising from laser-induced heating travels in the material at the longitudinal sound velocity \cite{Thomsen1986}.
In our experimental geometry, where the pumped area and the sample thickness are much larger than the pump and probe penetration depths, the relevant length scale for quantifying the strain-induced response is the smallest of these two depths.
Given a $v_s \sim$ 5400 m/s sound velocity at $T_f \sim$ 156 - 177 K -- an average between experimental values obtained in similar materials, GdMnO$_3$ and DyMnO$_3$ \cite{Lalitha2008} -- and a 76 nm probe penetration depth, we expect strain-induced signatures, averaged over the probe depth, to arise after about $\sim$14 ps.
This value is smaller than but comparable to the timescales seen in Fig. \ref{fig3}a and quantified in more detail in Table \ref{table}.
The slight difference in timescales can potentially be explained by anisotropy in the longitudinal sound velocity, which is not accounted for in the values above, obtained on powder samples, and by the fact that the estimate is made based on sound velocity values measured not on TbMnO$_3$ but on similar rare earth manganites.

As just discussed, photoexcitation leads to heating ($\Delta T \sim$ 136 - 157 K) and tensile strain ($\eta_b(t)>0$, Fig. \ref{fig3}a)) so that both must be considered when analyzing its effect on the physics of TbMnO$_3$.
We turn to previous reports that examine the dependence of the structural parameters on temperature \cite{Blasco2000}.
Figure \ref{fig3}b shows the relative change in lattice spacing (strain) along each axis and in the unit cell volume upon increasing the temperature from 20 K to RT, from Blasco \emph{et al.} \cite{Blasco2000}.
It is clear that the unit cell volume increases nearly monotonically as a function of temperature.
The lattice constants, however, exhibit a more complicated temperature dependence.
In particular, the value of $a$ decreases with temperature until about 210 K.
A direct comparison between the temperature dependent strain data (Fig. \ref{fig3}b) and the transient strain we measure (Fig. \ref{fig3}a) is hindered by the fact that, in the out-of-equilibrium case, the in-plane strain develops much more slowly than the strain normal to the surface.
In more detail, the stress created by heating, and in particular by heating from the laser, is a diagonal tensor, $\sigma_{ij}^T = a_{ii} \delta_{ij}$.
In the time resolved experiment, since we are looking at times $<< D_{laser}/v_s$, where $D_{laser} \sim 500$ $\mu$m is the lateral laser spot size and $v_s \sim$ 5400 m/s is the sound velocity reported above, there is insufficient time for the lattice strain along the in-plane $a$ and $c$ axes to change.
We therefore expect a strain only along $b$, which is related to the stress via
\begin{equation}
\sigma_{22}^{tot} = C_{2222} \: \eta_{22} -\sigma_{22}^T =0,
\label{straineq}
\end{equation}
where $C_{ijkl}$ is a generic element of the elasticity tensor of TbMnO$_3$.
This differs from the case when we change temperature in thermal equilibrium, for which the other components of the strain tensor can also react, resulting in additional contributions involving $C_{2211}$ and $C_{2233}$.
Using a reasonable estimate for $C_{ijkl}$ \cite{Hazama2000} we can solve Eq. \ref{straineq} for $\eta_{22}$ assuming that the system is heated by $\Delta T \sim$ 136 - 157 K.
The $b$-axis strain calculated in this way is, however, essentially the same as the one reported in Fig. \ref{fig3}b.
This is probably related to the fact that the strain along the $a$- and $c$-axis (the in-plane components in our $b$-cut samples) vary in opposite directions with increasing temperature, so that the average in-plane strain is small even in thermal equilibrium.
We can finally compare the transient $\eta_b(100$ ps) = 0.02\% value obtained with 9 mJ/cm$^2$ absorbed fluence and an estimated $T_f \sim$ 156 - 177 K with the $b$-strain value of 0.31 - 0.34\% obtained from Fig. \ref{fig3}b in that temperature range, and conclude that there is a factor of over an order of magnitude difference.
A 0.02\% $b$-strain is obtained at $T$ = 59 K instead (Fig. \ref{fig3}b), which would correspond to $\Delta T$ = 39 K and a ten times lower absorbed fluence.
These observations are consistent with those from magnetic order dynamics, mentioned above, where a $\sim$33 K spin temperature increase was determined following photoexcitation with a 9 mJ/cm$^2$ absorbed fluence \cite{Johnson2015}.\\


To better understand the discrepancy between the measured and expected values of the strain we have considered several possibilities.
Potential technical issues include an incorrect estimate of the absorbed fluence and of the deposited energy density.
Regarding the accuracy of our determination of the absorbed fluence, in addition to the TbMnO$_3$ samples we measured the structural response of bulk [411]-cut bismuth samples and found the data to be comparable to those from published reports \cite{Johnson2008}, and certainly not inaccurate by a factor of 10.
Calculating the energy density from the absorbed fluence requires knowing the penetration depth.
As detailed above, we performed a careful estimate based on the optical conductivity data from Baldini \emph{et al.} \cite{Baldini2018}, taking into account the anisotropy in the material.
Using other available optical conductivity data \cite{Bastjan2008}, obtained on samples with cuts different from ours, only leads to an enhancement of the discrepancy we observe.
We can therefore rule out that the low transient strain values are due to inaccuracies in the determination of the experimental parameters.

We now consider possible explanations for the low values measured for $\eta_b(t)$ arising from the physical properties of the material.
Summarizing the discussions above, 1.55 eV photons promote intersite $d-d$ transitions between Mn$^{3+}$ ions \cite{Bastjan2008, Moskvin2010}.
This electronic energy is then transferred to the lattice in the form of heat, typically in about a picosecond \cite{Wall2009, Baldini2018}.
Heating the lattice creates expansive thermal stress near the surface, leading to expansive uniaxial strain due to coherent longitudinal acoustic phonons \cite{Thomsen1986}.
Given that the transient strain we measure is over an order of magnitude smaller than that expected from the absorbed energy, there must be a bottleneck in one of these steps.
Starting with the last step, it is difficult to imagine a process by which the lattice temperature would increase but the generated strain would be much smaller than expected.
One possibility would be that heat conductivity is very high, such that the lattice cools down before strain has time to arise, but this is not expected in insulating TbMnO$_3$ \cite{Berggold2007}.
The most likely scenario is that there is a bottleneck in the transfer of energy between the electronic and lattice subsystems.
Within this scenario, one hypothesis would be that there is extremely weak coupling between the electron and lattice subsystems, i.e. an extremely low electron-phonon coupling constant (generally challenging to determine experimentally), which would extend the typical one picosecond timescale to beyond our 100 ps measurement window.
This would be quite unexpected, and would hinder the production of strain waves, which relies on heating-induced stress being created at the surface on a timescale faster than that of acoustic wave propagation \cite{Thomsen1986}.
The small strain we measure could, however, still arise from direct coupling of excited electronic states to long-wavelength acoustic modes via deformation of the ionic potential, without any contribution from lattice heating.
Theoretical modeling would be required to validate this hypothesis.
An alternative hypothesis is that a large fraction of the electronic excitations gets trapped and does not immediately contribute to lattice heating.
With such reduced lattice heating, the strain wave amplitude \cite{Thomsen1986} would also decrease, consistent with the small value of $\eta_b(t)$ that we measure.
One possible trapping mechanism that was proposed relies on the formation of anti Jahn-Teller polarons \cite{Allen1999, Talbayev2015, Johnson2015, Baldini2018}.
According to this mechanism, the photoexcited Mn$^{2+}$ - Mn$^{4+}$ pair leads to a local relaxation of the Jahn-Teller distortion which prevents further hopping.
The lifetime for this charge localization could well exceed the 100 ps pump-probe delay time accessible through our measurements.
A direct test of this second hypothesis would be to use a technique sensitive to polaron formation, such as x-ray diffuse scattering or pair distribution function (for smaller polaron sizes).
Independently of the origin of the bottleneck in the transfer of energy between the electronic and lattice subsystems, this scenario could in principle be confirmed by measuring the structural response of TbMnO$_3$ well beyond 100 ps and seeing whether the lattice slowly expands over time, or by using a technique sensitive to the lattice temperature such as transient Raman scattering.
A lattice heating process that is too gradual may, however, be compensated by heat diffusion and therefore be difficult to detect.

As a final discussion topic, we address the potential relationship between magnetic and structural dynamics.
Specifically, we are interested in finding out whether coupling to the lattice could be responsible for the relatively long 20 - 30 ps demagnetization timescale \cite{Handayani2013, Johnson2015, Bowlan2016, Bothschafter2017, Baldini2018}.
The magnetic order in TbMnO$_3$ is dictated by the superexchange interaction $J$ between Mn$^{3+}$ ions, bridged by O$^{2-}$ ions.
$J$ can be modified by changing
(i) the valence of the Mn$^{3+}$ ions or
(ii) the relative position of the Mn$^{3+}$ and O$^{2-}$ ions \cite{Fedorova2018}.
A change in ionic position can be achieved directly via strain or indirectly via heating, which leads to disorder and eventually to a change in the average position of the ions.
Regarding (i), photoexcitation at 1.55 eV creates an Mn$^{2+}$ - Mn$^{4+}$ pair from two Mn$^{3+}$ ions, meaning that it locally alters the valence of the Mn$^{3+}$ ions.
This process occurs in $<1$ ps \cite{Qi2012, Handayani2013, Baldini2018} but is seen to not lead to a large enough disorder in the spin system for the long range magnetic order to be affected on this short timescale \cite{Johnson2015, Bothschafter2017}, contrary to what is observed in e.g. antiferromagnetic CuO \cite{Johnson2012} and ferrimagnetic compounds \cite{Ogasawara2005}.
Trapped anti Jahn-Teller polarons have been suggested as an explanation for the 20 - 30 ps demagnetization time \cite{Allen1999, Talbayev2015, Johnson2015, Baldini2018}, the picture being that hopping is restricted until the lattice expands due to heating.
We have observed, however, that the lattice heats up much less than expected up to 100 ps ($\Delta T$ = 39 K instead of $\Delta T >$ 136 K), where the dynamics appear to saturate, a result that is confirmed by demagnetization dynamics data from Johnson \emph{et al.} \cite{Johnson2015} ($\Delta T$ = 33 K), as discussed above.
If anti Jahn-Teller polarons do form, we therefore expect them to be more long lived than the 20 - 30 ps demagnetization timescale.

An alternative explanation for the demagnetization is related to (ii), where a change in ionic position causes a change in $J$ by changing the hybridization between Mn$^{3+}$ and O$^{2-}$ ions \cite{Fedorova2018, Kimel2002, Ogasawara2005, Thielemann2017, Afanasiev2021}.
Our measurements provide a direct estimate of $\eta_b(t)$ and can be summarized as follows.
First, the value of $\eta_b(t)$ created by photoexcitation (Fig. \ref{fig3}a)) is smaller than expected but nevertheless similar to the $b$-axis strain which accompanies a loss of magnetic order in the system (i.e. corresponding to a lattice temperature larger than $T_{N_1}$, Fig. \ref{fig3}b)).
Second, strain propagates in TbMnO$_3$ on the same 20 - 30 ps timescale (Table \ref{table}) as is observed for demagnetization.
Third, the lattice heating which creates $\eta_b(t)$ (if any) necessarily occurs on a timescale faster than that of $\eta_b(t)$, although we could not measure the heating timescale directly.
Based on these observations, we cannot unequivocally distinguish between direct strain and heating-induced disorder as the driving force behind the changes in the average ionic position which are responsible for demagnetization.
However, independently of the details of the process, the fact that the demagnetization timescale coincides with the lattice dynamics and is over an order of magnitude slower than the electronic excitation enables us to conclude that the magnetic order in TbMnO$_3$ is robust to changes in the valence of the Mn$^{3+}$ ions and controlled by changes in the position of the Mn$^{3+}$ or O$^{2-}$ ions.\\


We have investigated the structural dynamics in the multiferroic phase of TbMnO$_3$ and have observed the appearance of tensile strain along the $b$-axis, perpendicular to the sample surface, following photoexcitation at 1.55 eV.
The $\sim$0.02\% tensile strain at 100 ps is over an order of magnitude smaller than the $>$ 0.31\% strain that arises in the system upon heating from 20 K up to the expected final temperature after photoexcitation, $T_f >$ 156 K.
We attribute this discrepancy to a bottleneck in the energy transfer between the electronic and lattice subsystems, possibly related to an extremely low electron-phonon coupling constant or to the formation of anti Jahn-Teller polarons.
Furthermore, the strain arises on the same 20 - 30 ps timescale that was reported for demagnetization, suggesting that the exchange interaction that governs magnetic order in TbMnO$_3$, which remains robust against local variations in Mn$^{3+}$ ionic valence, is stabilized by the lattice structure.
Our conclusions contribute to building a general phase diagram of rare-earth manganites, which is essential for our physical understanding of these systems as a whole, as well as for including them in technological devices in the future.\\


The research leading to these results has received funding from the Swiss National Science Foundation and its National Center of Competence in Research, Molecular Ultrafast Science and Technology (NCCR MUST) \emph{and Materials' Revolution: Computational Design and Discovery of Novel Materials (NCCR MARVEL).}
E. A. acknowledges support from the ETH Zurich Postdoctoral Fellowship Program and from the Marie Curie Actions for People COFUND Program, as well as from the Swiss National Science Foundation through Ambizione Grant PZ00P2\_179691.
We acknowledge the Paul Scherrer Institute (PSI) for synchrotron radiation beamtime.
We thank Dr. E. Pomjakushina at PSI for help with sample preparation, as well as the MicroXAS beamline staff for assistance, in particular Alexander R. Oggenfuss.


\bibliographystyle{apsrev4-1}
\bibliography{TbMnO3FEMTO}

\end{document}